\documentclass{article}
\usepackage{graphicx}
\usepackage{times}
\begin{document}
\begin{titlepage}
\title{Differential rotation and meridional flow of Arcturus}


\author{
          M.~K\"uker, G.~R\"udiger
\\[5mm]
%
   Astrophysikalisches Institut Potsdam (AIP)\\
   An der Sternwarte 16, D-14482 Potsdam  \\
   email: mkueker@aip.de, gruediger@aip.de
   }
\end{titlepage}
\maketitle
\abstract
{The spectroscopic variability of Arcturus hints at cyclic activity cycle and differential rotation. This could provide a test of current theoretical models of solar and stellar dynamos. To examine the applicability of current models of the flux transport dynamo to Arcturus, we compute a mean-field model for its internal rotation, meridional flow, and convective heat transport in the convective envelope. We then compare the conditions for dynamo action with those on the Sun.
We find solar-type surface rotation with about 1/10th of the shear found on the solar surface. The rotation rate increases monotonically with depth at all latitudes throughout the whole convection zone.  In the lower part of the convection zone the horizontal shear vanishes and there is a strong radial gradient. The surface meridional flow has maximum speed of 110 m/s and is directed towards the equator at high and towards the poles at low latitudes. Turbulent magnetic diffusivity is of the order $10^{15}$--$10^{16}  {\rm cm^2/s}$. The conditions on Arcturus are not favorable for a circulation-dominated dynamo.}
%
%

\section{Introduction}
Arcturus is a single giant star of spectral type K2 with an effective temperature of 4300K (Griffin \& Lynas-Gray 1999). Gray \& Brown (2006) found a two year modulation in the velocity span of the bisector of the Fe I $\lambda$6252.57 line which they interpret as the rotation period of the star. The two year period has also been found in the Ca II emission of Arcturus which has been monitored by the Mount Wilson H+K project since 1984 (Brown et al.~2008). The data also shows a variation on a longer time scale that could be an activity cycle with a period $\le$ 14 years. The so observed rotation period varies with an amplitude of 70days, changing by 20 days/year. A similar time dependence is found in the solar Ca II H+K emission, where it is a consequence of the differential rotation. As the activity moves to lower latitudes during the cycle, the rotation period decreases. By analogy, the variation in the rotation period of Arcturus can be interpreted as a variation of the active latitude on a differentially rotating stellar surface. Stellar butterfly diagrams are ambiguous, however, as a combination of anti-solar rotation and a polewards drift of the active latitude will produce the same pattern as if dynamo and rotation were both solar-type.

The solar butterfly diagram can be explained as the result of an $\alpha \Omega$ dynamo, i.e.~a combination of differential rotation and the $\alpha$ effect caused by the helicity of the convective gas motions. However, significant radial shear is only found in the subsurface layer and in the tachocline at the bottom of the convection zone. More recent dynamo models therefore explain the butterfly diagram with the advection of magnetic flux by the large-scale meridional flow (cf.~Ossendrijver 2003, Charbonneau 2005). 

Flux transport by the meridional flow is only effective if the associated magnetic Reynolds number is large, i.e.~
\begin{equation}
  {\rm Rm} = \frac{u^{\rm m} d}{\eta} \gg 1
\end{equation}  
where $u^{\rm m}$ is the meridional flow speed, $d$ a characteristic length scale (e.g.~the stellar radius) and $\eta$ the magnetic diffusivity coefficient (K\"uker et al.~ 2001). 

The solar differential rotation is the result of angular momentum transport by convection and the large-scale meridional flow in the convection zone. On the one hand stratification causes a non-diffusive contribution to the Reynolds stress in addition to the diffusive part known as turbulence viscosity, on the other hand the Coriolis force causes a deviation of the convective heat flux from the radial direction. The result is a small horizontal temperature gradient that drives a meridional circulation. Mean field models of the solar convection zone reproduce the observed rotation pattern and surface meridional flow very well and allow predictions for other stars (Kitchatinov \& R\"udiger 1995, 1999, K\"uker \& Stix 2001).

To examine the conditions for dynamo action on Arcturus, we apply our stellar rotation model to its convective envelope. Like solar-type stars, giants have deep outer convection zones. One might therefore expect similar rotation and magnetic activity patterns. There are, however, significant differences. With an equatorial rotation period of 25 days the Sun rotates much faster than Arcturus and with an effective temperature of 5780 K it is substantially hotter. Interestingly, the radius of Arcturus is larger by about the same factor by which its rotation period is longer, yielding about the same equatorial rotation speed (1.8 km/s vs.~2km/s).  Finally, the geometrical depth of the outer convection zone is much larger than that of the Sun, both in absolute values and relative to the stellar radius. 
In the model described below it reaches down to three percent of the stellar radius, making the star almost fully-convective. 
%
%
\section{The model}
%
%
\subsection{Angular momentum transport}
We describe the convective motions in the star with the mean-field ansatz, 
\begin{equation}
  \vec{u}=\vec{\bar{u}}+\vec{u'}
\end{equation}
where $ \vec{\bar{u}}$ is the average and $\vec{u'}$ the fluctuating part. Assuming that the mean gas flow to be stationary, it is governed by the Reynolds equation,
\begin{equation} \label{reynolds}
  \rho (\vec{\bar{u}}\cdot \nabla)
       \vec{\bar{u}}  =  - \nabla \cdot \rho Q
	   - \nabla P + \rho \vec{g},
\end{equation}
where 
$
   Q_{ij} =  \langle u_i' u_j' \rangle
$
is the one-point correlation tensor of the velocity fluctuations. Together with the density it constitutes the Reynolds stress:
\begin{equation}
  {\cal T}_{ij} = - \rho Q_{ij}.
\end{equation}
  
With the assumption of axisymmetry the azimuthal component of Eq.\ref{reynolds} becomes a conservation equation for angular momentum:
\begin{equation} \label{omega}
    \nabla \cdot \vec{t} = 0,
\end{equation}
with the flux vector
\begin{equation}
    \vec{t} =  r \sin \theta \left [ \rho r \sin \theta
    \Omega \vec{\bar{u}}^{\rm m}
    + \rho \langle u_{\phi}' \vec{u}' \rangle \right],
\end{equation}
where $\vec{\bar{u}}^{\rm m}$ is the large-scale meridional flow.

For the heat transport the mean-field ansatz leads to the equation
\begin{equation} \label{heat}
    \nabla \cdot  (\vec{F}^{\rm conv} + \vec{F}^{\rm rad}) 
= 0,
\end{equation}
where 
$\vec{F}^{\rm conv} = \rho \langle \vec{u'} s' \rangle$, with the specific entropy $s$, is the convective heat flux and $\vec{F}^{\rm rad}$ the radiative heat flux.

Equations \ref{reynolds} and \ref{heat} are solved in spherical coordinates. As the fluctuations enter only through their correlations $\langle u'_i u'_j \rangle$ and $\langle \vec{u'} s' \rangle$ no detailed  knowledge of the small-scale motions is required. Analytical expressions for the correlations can be derived using the Second Order Correlation Approximation (K\"uker et al.~1993, Kitchatinov et al.~1994, Kitchatinov \& R\"udiger 2005). 

The convective heat flux then reads
\begin{equation} \label{heatflx}
  F^{\rm conv}_i =-\rho T \chi_t \Phi_{ij} \frac{\partial s}{\partial x_j},
\end{equation}
where the dimensionless coefficients $\Phi_{ij} \le 1$ are functions of the Coriolis number,
 \begin{equation}
 \Omega^*=2 \tau_c \Omega,
\end{equation} 
where $\tau_c$ is the convective turnover time and $\Omega$ the angular velocity of the stellar rotation. For slow rotation Eq.~\ref{heatflx} is identical with the corresponding expression from standard mixing-length theory but for fast rotation ($\Omega^*$ not small) the efficiency of the convective heat transport is reduced and the corresponding flux vector tilted towards the rotation axis.
   
The Reynolds stress has a diffusive (turbulence viscosity) and a non-diffusive ($\Lambda$ effect) part:
\begin{equation}
  Q_{ij} = - \nu_{ijkl} \frac{\partial u_k}{\partial x_l} + \Lambda_{ijk} \Omega_k.
\end{equation}
Like the heat transport, the stress takes the form of standard isotropic diffusion in case of slow rotation, i.e.
\begin{equation}
  Q_{ij} \approx - \nu_t (\frac{\partial u_i}{\partial x_j}+\frac{\partial u_j}{\partial x_i}) 
   \hspace{0.5cm} {\rm for} \hspace{0.5cm}\Omega^* \ll 1.
\end{equation}
For fast rotation ($\Omega^* \ge 1$) the stress is not only anisotropic but actively builds up a gradient in the angular velocity as rigid rotation is not stress-free because of the $\Lambda$ effect. 

Note that the convective turnover time is depth-dependent, which implies a depth-dependence of the Coriolis number. The lower parts of a stellar convection zone with its longer time scale is thus more affected by the stellar rotation that the upper part with its shorter time scale.

We apply the numerical scheme described in K\"uker et al.~(2010). As boundary conditions we assume stress-free and impenetrable boundaries for the gas motion and an imposed radial heat flux. At the lower boundary the heat flux is constant: corresponding to the stellar luminosity for the heat transport,
   \begin{equation}
      \vec{F}_r = \frac{L}{4 \pi r_b^2}
   \end{equation}
   where $L$ is the stellar luminosity and $r_b$ the radius at which the boundary is located. At the upper boundary the heat flux is allowed to vary with latitude:
   \begin{equation}
    \vec{F}_r = \frac{L}{4 \pi r_t^2} \left(1+\frac{\delta T}{T}\right)^4
     \approx \frac{L}{4 \pi r_t^2} \left(1+\frac{4 \delta s}{c_p}\right),
   \end{equation}
where $r_t$ is the location of the upper boundary, $c_p$ the specific heat capacity, and $\delta T$ and $\delta s$ are the deviations of the temperature and entropy, respectively,  from their corresponding values for strictly adiabatic stratification.

   As the radiative heat flux is prescribed, we define the bottom of the convection zone as the point where the radiative heat flux equals the total heat flux. This condition is equivalent to imposing zero convective flux there. The boundary conditions on the rotation axis are implied by axisymmetry.
\subsection{Model star}
While the radius of Arcturus can be inferred from its interferometric diameter and the Hipparcos parallax (Gray \& Brown 2006), a reliable value for its mass has so far not been determined (Trimble \& Bell 1981, Griffin \& Lynas-Gray 1999). We thus have some freedom in the choice of our model star.
%
%
 The metal abundaces of Arcturus differs substantially from that of the Sun (Worley et al.~2005). However, a model from an evolutionary track for a $1.5 M_\odot$ star and solar abundances computed with the Mesa code\footnote{http://mesa.sourceforge.net} is reasonably close enough to the observations for our purposes. (See the comparison with the alternative model below). It has an effective temperature of 4170 K, 27 solar radii and a radiative core of  3\% of the stellar radius. For the purpose of computing the differential rotation, the convection zone is assumed to be in hydrostatic equilibrium and the stratification near-adiabatic. 
\begin{figure}
 \begin{center}
   \includegraphics[width=4.1cm]{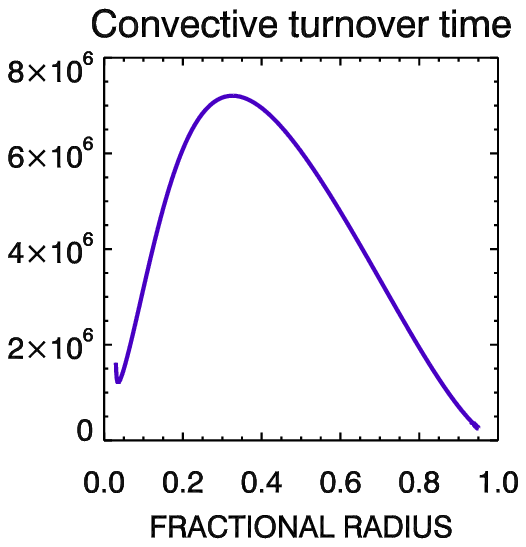}
   \includegraphics[width=4.1cm]{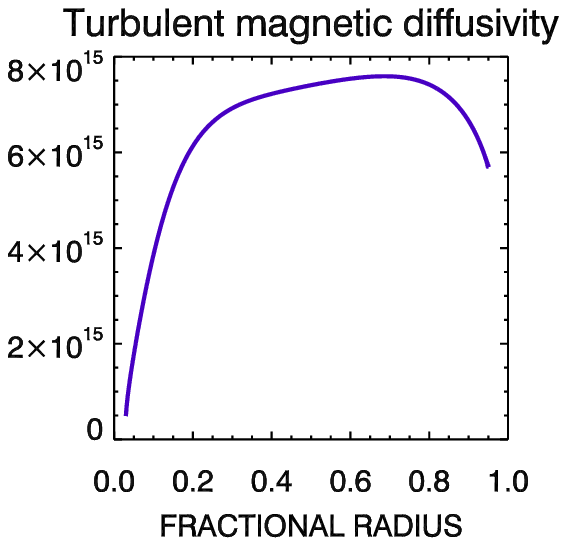}
 \end{center}
 \caption{\label{etaturb}
       Left: Convective turnover time in s vs. fractional radius. Right: Turbulent magnetic diffusion coefficient in cm$^2$/s vs. fractional radius. }
\end{figure}

The left panel of Fig.~\ref{etaturb} shows the convective turnover time as a function of the fractional radius. The large stellar radius implies a low value of the surface gravity of log g=1.7 and corresponding values throughout the convection zone. The resulting convective time scale of 1-2 months  leads to a Coriolis number about unity, which is half the solar value despite a rotation period much longer than that of the Sun. The right panel shows the turbulent magnetic diffusion coefficient, 
\begin{equation}
  \eta_t=\frac{1}{3} l_{\rm mix} u_c, 
\end{equation} 
where $l_{\rm mix}$ is the mixing length and $u_c$ the convection velocity. The values of 6--8$\times 10^{15}$cm$^2$/s in the bulk of the convective envelope are 2--3 orders of magnitude larger than the corresponding value in the solar convection zone. 
%
\section{Results}
\begin{figure*}
   \begin{center}
   \includegraphics[width=12cm]{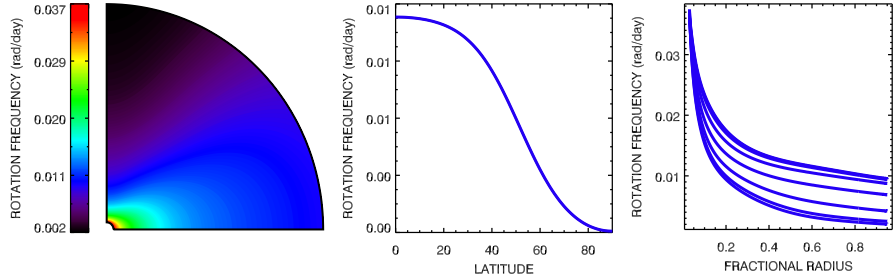}
   \end{center}
   \caption{ \label{profile}
   Left: colour contour plot of the internal rotation rate. Center: rotation rate as a function of latitude at the upper boundary. Right: rotation rate as function of fractional radius for 0 (equator), 15, 30, 45, 60, 75, and 90 (poles) degree latitude, from top to bottom.
   }
\end{figure*}

%
Solving the equations of motion and heat transport for the convection zone described above and a rotation period of two years yields the rotation pattern shown in Figure \ref{profile}. The surface rotation is solar-type, e.g. the rotation period is shortest at the equator and longest in the polar caps. The surface shear,
\begin{equation}
  \delta \Omega = \Omega_{\rm eq} - \Omega_{\rm pol},
\end{equation}
amounts to 0.008 rad/day. This is only marginally less than the rotation rate at the equator as the latter is an order of magnitude larger than the polar rotation rate. This means that while the total shear is smaller than on the solar surface, the relative shear, $\delta \Omega / \Omega_{\rm eq},$ is close to one. Overall the surface shear is about an order of magnitude smaller than that of the Sun, which shows a difference of 0.065 rad/day between the angular velocities at the equator and 75 deg latitude (cf.~Miesch 2005).

While the horizontal shear dominates in the outer layers, the rotation rate is constant with latitude but varies strongly with radius at the bottom of the convection zone. Throughout the whole region, the rotation rate decreases with increasing radius.
\begin{figure*}
  \begin{center}
   \includegraphics[width=12.0cm]{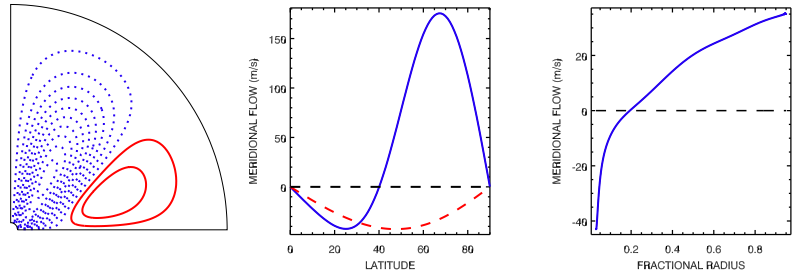}
  \end{center} 
 \caption{ \label{flow}
 Left: Stream lines of the meridional flow. Solid red lines denote clockwise circulation, dotted blue lines counterclockwise. Center: The meridional flow at the top (blue line) and bottom (red line) of the convection zone. Right: the meridional flow speed as a function of the fractional stellar radius at 45 deg latitude. Positive values mean that the gas flow is towards the equator, negative values indicate gas motion towards the pole. 
 }  
\end{figure*}

Figure \ref{flow} shows the  meridional flow pattern. There are two flow cells per hemisphere, a cell of slow flow at low latitudes and a cell of much faster flow at mid-high latitudes. The orientation of the dominant flow cell is opposite to that of the solar meridional flow, i.e. the surface flow is directed towards the equator. The maximum flow speed of 170 m/s is reached at the surface at mid-high latitudes.

The flow pattern depends on the rotation period. For slow rotation the flow is mainly driven by differential rotation while for fast rotation the baroclinic force dominates. At the observed period of two years the two effects are of equal strength. The baroclinic force dominates at high latitudes, the differential rotation at the equator. 

\begin{figure}
  \begin{center}
  \includegraphics[width=9.0cm]{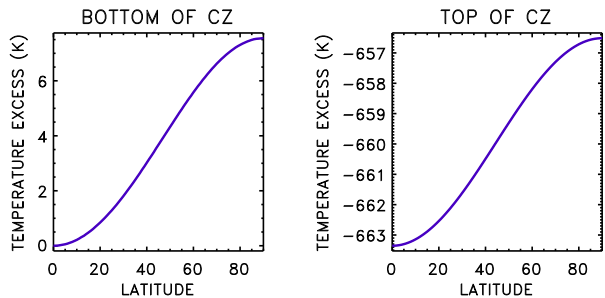}
  \end{center}
  \caption{ \label{tsup}
    Temperature deviation $\delta T$ at the bottom (left) and top (right) of the convection zone.
    }
\end{figure}
Figure \ref{tsup} shows the horizontal variation of the temperature resulting from the tilt of the convective flux. Though the difference between the polar caps and the equator is only about 7 K, it has a profound effect on the meridional flow and (indirectly) on the differential rotation.
%
%
\subsection{Alternative model}
\begin{figure}
\begin{center}
\includegraphics[width=8.0cm]{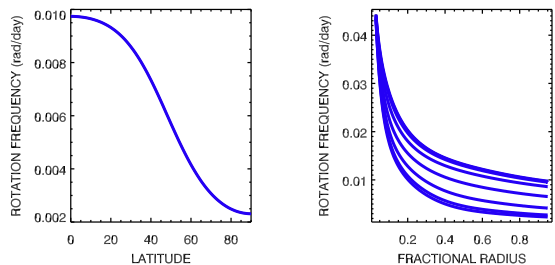}
\end{center}
\caption{ \label{m10rot}
Differential rotation of the 1 $M_\odot$ model. Left: rotation rate vs.~latitude. 
 Right: rotation rate vs.~radius for the same latitudes as in Fig.~\ref{profile}. 
 }
 \end{figure}

\begin{figure}
  \begin{center}
   \includegraphics[width=8.0cm]{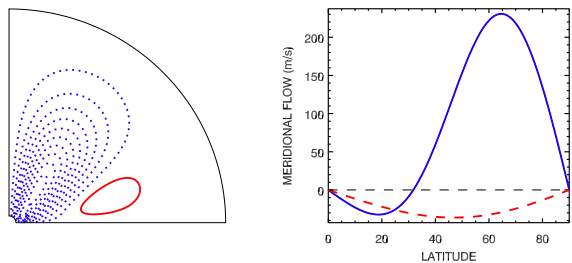}
  \end{center} 
 \caption{ \label{m10flow}
 Meridional flow of the one solar mass star. 
 Left: stream lines. The solid red line indicates clockwise circulation, the doted blue lines counter-clockwise circulation. Right: Horizontal flow speed at the top (solid blue line) and bottom (dashed red line) of the convection zone. 
 }  
\end{figure}

As the mass of Arcturus is not very well determined we repeat our computation with a model from an evolutionary track for a star of one solar mass and $Z=0.005$. The resulting rotation and meridional flow patterns are shown in Figs.~\ref{m10rot} and \ref{m10flow}. The rotation pattern is very similar to the 1.5 $M_\odot$ model. The surface rotation shows the same shear of $\approx \delta \Omega$ = 0.008 rad/day and the pole rotates with an angular frequency of 0.002 rad/day. The lower boundary rotates faster (0.045 rad/day vs.~0.037 rad/day) and the radial shear is thus more pronounced than with the 1.5 $M_\odot$ mode. The meridional flow shows a similar two-cell pattern but the low-latitude cell is even weaker. The maximum flow speed is slightly higher than for the 1.5 $M_\odot$ model, reaching 230 m/s rather than 170 m/s. 
%
 
%
%
\section{Discussion}
%
%
Despite the great geometrical depth, the properties of the convection zone of Arcturus are closer to those of the solar granulation and supergranulation layers than the deeper parts of the solar convection zone because of the long rotation period. With two years the latter is an order of magnitude longer than the convective turnover time, which reaches a maximum value of 85 days at a fractional radius of 0.4. Except for the layer around that radius the Coriolis number is less than one, making Arcturus a slow rotator in this context. 

The rotation pattern shows a strong variation with depth. For fractional radii greater than 0.2 there is a moderate decrease of the rotation rate with increasing radius and "solar-type" horizontal shear, i.e. the rotation rate is larger at the equator than at the poles. While the latter difference is large in relative terms it is only about 1/10th of the solar surface differential rotation in absolute terms. The most striking feature is the steep decline of the rotation rate with increasing radius at the lower boundary. Such a rotation profile would imply a core that rotates much faster than the convective envelope. 

A rotation rate that decreases with increasing radius is the natural consequence of angular momentum transport by Reynolds stress. For slow rotation the radial transport dominates and creates a negative gradient in the rotation rate, as is observed in the outermost layers of the solar convection zone.
 
The meridional flow is much different from that found for the Sun. Both observations and mean field models find a surface flow of about 20 m/s amplitude towards the poles. The return flow has not been observed yet but is predicted by theory to occur at the bottom of the convection zone with about half the amplitude of the surface flow. Our model for Arcturus shows two flow-cells per hemisphere rather than a solar-like one-cell pattern. There is a large cell of fast flow at high latitudes and a smaller cell with lower flow speeds at low latitudes. The high-latitude flow is anti-solar, i.e. directed towards the equator at the surface while the slower flow at low latitude is solar-type.   

The situation for giant stars such as Arcturus is much different from that for Main Sequence stars as for the latter the mass is fixed by the effective temperature. For single giants there is a much larger uncertainty. Our alternative 1 $M_\odot$ model shows, however, that the differential rotation does not depend very strongly on the stellar mass provided temperature and radius are kept constant. 
 
Our model does not include the stably stratified core. In the Sun, the surface differential rotation persists throughout the whole convection zone while the core rotates rigidly with the same rotation rate as the surface at mid-latitudes. The transition between the two patterns occurs in a shallow layer at the bottom of the convection zone, the tachocline. It has been the subject of much debate both about the mechanism that causes such a sharp transition and the role it may play in the solar dynamo.  

With Arcturus we find a different situation. The horizontal shear ($\partial \Omega/\partial \theta$) disappears in the lower part of the convective envelope and a strong radial gradient appears. As the rotation is (horizontally) uniform at the lower boundary, a solar-type tachocline can not exist.
The sharp decrease of the rotation rate with radius raises the question whether this rotation profile is dynamically stable. We can exclude the Taylor-Couette instability as the rotation rate decreases roughly like $1/\sqrt{\Omega}$, implying hydrodynamic stability. However, MHD instability caused by a sufficiently strong toroidal field can not be ruled out without a detailed dynamo model. 

The rotation pattern we find for Arcturus looks similar to the one assumed by early models of the solar dynamo. The combination of negative radial shear and the positive (negative) $\alpha$ effect that is expected in the northern (southern) hemisphere of a stellar convection zone (R\"udiger \& Kichatinov 1992) naturally produces a solar type butterfly diagram through a classical $\alpha \Omega$ dynamo without the need for a meridional circulation (Yoshimura 1975). This prompts us to have a look at the conditions for dynamo action.
 
As the star is almost fully convective and has no horizontal shear at the bottom of the convection zone, there is no tachocline. Moreover, the turbulent magnetic diffusivity computed from the mean field model is 2--3 orders of magnitude larger than the corresponding value for the solar convection zone, as shown in Fig.~\ref{etaturb}.  With meridional flow speeds of the order 100 m/s and a length scale of $2 \times 10^{12}$cm we find Rm$ \approx 2$. For the Sun a similar estimate yields a value of about 20, assuming a magnetic diffusivity coefficient of $10^{13}$ cm$^2$/s. Hence, conditions for a flux-transport dynamo are probably less favorable on Arcturus than on the Sun.

There is some uncertainty about the turbulent magnetic diffusivity in mean field dynamos. Current models of the solar dynamo use smaller values than predicted by the SOCA. Dikpati \& Charbonneau (1999) assume $10^{10}$--$10^{11}$ cm$^2$/s. Chatterjee et al.~(2004) use similar values for the toroidal field but larger values of the order $10^{12}$ cm$^2$/s for the poloidal field. 
To create similarly favorable conditions for the flux transport dynamo in Arcturus we have to lower the magnetic diffusivity by as much as three orders of magnitude to $10^{13}$ cm$^2$/s to reach magnetic Reynolds numbers of the order 1000, as required for this type of dynamo.   
\\[5mm]
{\it Acknowledgements:}
This work was supported by Deutsche Forschungsgemeinschaft.

\end{document}